\documentclass[twocolumn,showpacs,preprintnumbers,amsmath,amssymb]{revtex4}
\usepackage{graphicx}
\usepackage{dcolumn}
\usepackage{bm}

\newcommand{\ie}{{\it i.e.}}

\def\ave#1{\langle~#1~\rangle}

\begin{document}

\title{Spin dynamics of two-dimensional electrons with Rashba
spin-orbit coupling and electron-electron interactions}

\author{Yuan Li$^{1,2}$ and You-Quan Li$^1$\\
{$^1$Department of Physics,
Zhejiang University,  Hangzhou 310027, P. R. China\\
$^2$Institute of Materials Physics, Hangzhou Dianzi University,
Hangzhou 310018, P. R. China }}

\begin{abstract}
We study the spin dynamics of two dimensional electron gases (2DEGs)
with Rashba spin-orbit coupling by taking account of
electron-electron interactions. The diffusion equations for charge
and spin densities are derived by making use of the path-integral
approach and the quasiclassical Green's function. Analyzing the
effect of the interactions, we show that the spin-relaxation time
can be enhanced by the electron-electron interaction in the
ballistic regime.
\end{abstract}

\date{\today} \pacs{72.25.Rb, 71.70.Ej, 71.10.Ca}

\maketitle

\section{Introduction }

Spin based electronics~\cite{Wolf} or spintronics~\cite{Zutic} has
been an active research area in the past decade. The effort for
effectively manipulating electron spin by means of an applied
electric field~\cite{Datta,Schliemann,Murakami} is an important
issue there. The system with spin-orbit (SO) couplings makes those
efforts possible and thus brings great interests from both academic
and practical aspects recently. Thus, it is essential to study the
spin relaxation for further development of spintronics.

There are four main mechanisms of spin relaxation in
semiconductor systems
~\cite{Zutic,Pikus,Averkiev1999,Averkiev2002,Flatte,Li}.
In Elliott-Yafet mechanism, the spin-orbit coupling induces a mixing of
wave functions for valence-band states and conduction-band states.
The mixing that results in the spin relaxation of electrons is due
to the scattering by impurities or phonons.
The Elliott-Yafet
mechanism operates in semiconductors with and without a center of
inversion symmetry, while it is most prominent in the
centrosymmetric ones (such as silicon).
The Bir-Aronov-Pikus mechanism is applicable for $p$-doped semiconductors
in which the electron spin flipping is induced by exchange interaction with
holes. The hyperfine interaction provides another important
mechanism~\cite{Dyakonov1} for ensemble spin dephasing and single
spin decoherence of localized electrons.
The D'yakonov-Perel mechanism depicts that electrons can feel an effective random
magnetic field arising from the spin-orbit coupling in systems with
inversion asymmetry such that spin relaxation occurs.
This mechanics can interpret the spin dephasing in crystals without inversion
center and is particularly applicable for $n$-type samples.
For two dimensional $n$-type semiconductor systems without inversion
symmetry, the D'yakonov-Perel mechanism is believed to be most
important in wide ranges of carrier temperature and concentration.
Under certain conditions, the Elliott-Yafet mechanism may affect
spin dynamics of two-dimensional electrons in these systems.
The Bir-Aronov-Pikus mechanism is important for $p$-type semiconductor
systems and the hyperfine-interaction mechanism dominates for
localized electrons.

Most studies of spin relaxation in semiconductors have focused on
impurity (somewhat less phonon) mediated spin flips while neglecting
the effect of electron-electron interactions for a long time.
It has been noticed recently that electron-electron interactions
play certain role in spin relaxation and dephasing in semiconductor systems.
The electron-electron interaction is known to play a crucial role
in determining the transport and thermodynamic properties near the
metal-insulator transition in two-dimensional electron systems~\cite{Kravchenko},
which is suspected to affect the spin relaxation for the spin susceptibility
behaving critically when the metal-insulator transition
occur~\cite{Prus,Shashkin,Punnoose1}.
There are several experimental
and theoretical studies on the effect of electron-electron
interactions on spin relaxation.
The electron-electron scattering results in
additional momentum relaxation which induces spin
dephasing of electrons through the motional narrowing of the
D'yakonov-Perel type~\cite{Glazov1}
as measured in n-GaAs/AlGaAs quantum wells~\cite{Stich,Leyland}.
The electron-electron scattering effect on the spin
dephasing has been considered\cite{wu1} in a magnetic field,
and a momentum dependent
effective random magnetic field induced by the electron-electron
exchange interaction can lead to spin dephasing of
electrons~\cite{Weng1,Weng2,Glazov2}.
It is also observed that the
spin relaxation caused by the D'yakonov-Perel mechanism gives
considerably different rates depending on the technique
employed~\cite{Punnoose2}.

However, as we are aware, the explicit form of the diffusion
equation for two-dimensional electron gases (2DEGs) with spin-orbit
couplings has not been derived by taking account of
electron-electron interactions. It is thus obligatory to develop the
explicit form of the diffusion equation to study the spin dynamics
for 2DEGs with spin-orbit couplings as well as electron-electron
interactions. In this paper, we focus attention on the
D'yakonov-Perel spin-relaxation mechanism.
We investigate the spin
dynamics of electrons in two-dimensional $n$-type semiconductor
systems with electron-electron interactions and Rashba spin-orbit
coupling.

The paper is organized as follows. In Sec.~\ref{sec:phifield}, we
take account of the electron-electron interaction for the 2DEGs with
the Rashba spin-orbit coupling. Applying the path integral
formulation, we decoupled the interaction in terms of an auxiliary
Bose field. In Sec.~\ref{sec:kinetic}, we employ the quasiclassical
Green's function to investigate the spin dynamics of electrons. In
Sec.~\ref{sec:dynamics}, the  diffusion equations for spin and
charge densities as well as
 the explicit expression of spin-relaxation time are derived.
A summary is given in Sec.~\ref{sec:summary}
and some complicated formulae are given in the Appendix.

\section{Auxiliary fields describing the electron-electron interaction}
\label{sec:phifield}

Taking the electron-electron interaction into account, we study the
spin dynamics of electrons in two-dimensional systems with structure
inversion asymmetry. As the Fourier transform of the Coulomb
repulsion between electrons reads $V(\mathbf{q})=2\pi
e^2/|\mathbf{q}|$, the Hamiltonian of such a system is given by
\begin{eqnarray}\label{hamiltonian}
\hat{H}&=&\int \Big\{ \sum_{\lambda,\lambda'}
\hat{\psi}^{\lambda\dag}(r)
\big[\big(-\frac{\hbar^2}{2m}\nabla^2+U(r)-\mu\big)\delta_{\lambda,\lambda'}\nonumber\\
&&+\mathbf{b}\cdot
\vec{\sigma}_{\lambda\lambda'}\big]\hat{\psi}^{\lambda'}(r)\Big\}d^2r\\
\nonumber &&+\frac{1}{A}\sum_{\mathbf{q}\neq 0}\frac{\pi
e^2}{|\mathbf{q}|}\hat{\rho}(\mathbf{q})\hat{\rho}(-\mathbf{q}),
\end{eqnarray}
where $\hat{\psi}^{\lambda\dag}(r)$ and $\hat{\psi}^\lambda(r)$
represent the field operators with $\lambda=\uparrow,\downarrow$
labelling the spin state of the electron, $\hat{\rho}(\mathbf{q})$
represents the Fourier transform of the density operator
$\hat{\rho}(\mathbf{r})=\sum_\lambda
\hat{\psi}^{\lambda\dag}(r)\hat{\psi}^\lambda(r)$ and $\vec \sigma
=(\sigma_x, \sigma_y, \sigma_z)$ the Pauli matrices in spin space,
$U(r)$ a random disorder potential and $\mu$ the chemical potential.
The other notions in Eq.~(\ref{hamiltonian}) are $A=L^2$ with $L$
referring to the size of the sample and
$\mathbf{b}=\alpha\mathbf{p}\times \mathbf{e}_z$ with $\alpha$
referring to the Rashba spin-orbit coupling strength.
\begin{figure}[t]
\includegraphics[width=4.1cm]{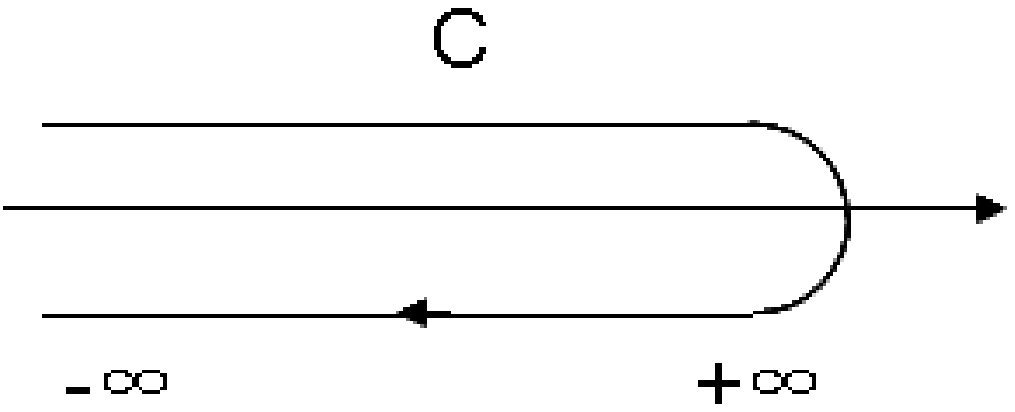}
\caption{\label{fig:contour} The Keldysh contour.}
\end{figure}
In holonomy representation (or the called coherent state
representation), the Green's function can be expressed as a
functional integral over the Grassmann fields $\psi^\lambda$ and
$\bar{\psi}^\lambda$ that reflect the fermionic nature of electrons,
\begin{eqnarray}\label{green1}
G_{\lambda\lambda'}(r,t;r',t')
=\ave{\psi^\lambda(r,t)\bar{\psi}^{\lambda'}(r',t')}\hspace{19mm}
    \nonumber\\
=\frac{\displaystyle\int D\bar{\psi}D\psi
       \psi^\lambda(r,t)\bar{\psi}^{\lambda'}(r',t')
        e^{-iS[\psi,\bar{\psi}]}
      }{\displaystyle\int D\psi D\bar{\psi}
      e^{-iS[\psi,\bar{\psi}]}
      }.
\end{eqnarray}
Here we adopted the unit $\hbar=1$
and the simplified notation $D\psi=D\psi^\uparrow D\psi^\downarrow$.
The action $S[\psi,\bar{\psi}]$ in the above equation is given by
\begin{eqnarray}\label{eq:action}
S[\psi,\bar{\psi}]&=&\int dt
 \Big\{\int d^2r\sum_{\lambda\lambda'}\bar{\psi}^\lambda(\mathbf{r},t)
  W_{\lambda\lambda'}\psi^{\lambda'}(\mathbf{r},t)\nonumber \\
&&+\frac{1}{A}\sum_{\mathbf{q}\neq 0}\frac{\pi
e^2}{|\mathbf{q}|}\big(\rho(\mathbf{q},t)\rho(-\mathbf{q},t)\big)\Big\},
\end{eqnarray}
where $W_{\lambda\lambda'}=\big(-i\partial/\partial
t-\nabla_r^2/2m+U(r)-\mu\big)\delta_{\lambda\lambda'}+\mathbf{b}\cdot
\vec{\sigma}_{\lambda\lambda'}$.

We divide the fermionic field
$\psi^{\lambda}(\mathbf{r},t)$ into two components
$\psi_{1}^{\lambda}(\mathbf{r},t)$ and
$\psi_{2}^{\lambda}(\mathbf{r},t)$ which reside,
respectively,
on the upper and lower branches of the Keldysh time contour shown in
Fig.~(\ref{fig:contour}).
Hence, the second line of Eq.~(\ref{eq:action}), which refers to
the interaction part, can be written as
$S_{int}[\psi_1, \bar{\psi}_1]-S_{int}[\psi_2,\bar{\psi}_2]$
with
\begin{eqnarray*}
S_{int}[\psi_i,\bar{\psi}_i]=\int dt \sum_{\mathbf{q}\neq
0}\frac{\pi
e^2}{A|\mathbf{q}|}\big(\rho_i(\mathbf{q},t)\rho_i(-\mathbf{q},t)\big),
   \\
\end{eqnarray*}
where $i=1, 2$. With the help of two auxiliary bosonic fields
$\tilde{\phi}_i(\mathbf{r},t)$, we can decouple those two terms
relevant to electron-electron interactions via the
Hubbard-Stratonovich transformation~\cite{Nagaosa}, namely,
\begin{eqnarray*}
&&\exp\big[-i\int dt \sum_{\mathbf{q}\neq 0}\frac{\pi
e^2}{A|\mathbf{q}|}\rho_i(\mathbf{q},t)\rho_i(-\mathbf{q},t)\big]\nonumber\\
&=&\int D\tilde{\phi}_i(\mathbf{q},t)\exp\big[i\int dt
\sum_{\mathbf{q}\neq
0}\frac{|\mathbf{q}|}{{4\pi}}\tilde{\phi}_i(\mathbf{q},t)\tilde{\phi}_i(-\mathbf{q},t)\big]\nonumber\\
&&\times \exp\Big[i\int dt \frac{e}{2\sqrt{A}}\sum_{\mathbf{q}\neq
0}\big\{\tilde{\phi}_i(\mathbf{q},t)\rho_i(-\mathbf{q},t)
\nonumber\\&&+\rho_i(\mathbf{q},t)\tilde{\phi}_i(-\mathbf{q},t)\big\}
\Big].
\end{eqnarray*}
Then we can write the Green's function as follows,
\begin{eqnarray}\label{eq:Green3}
&&\hat{G}_{\lambda\lambda'}(r,t;r',t')
  \nonumber\\
&&=\frac{\displaystyle\int D\bar{\Psi}D\Psi
D\Phi \Psi^\lambda(r,t)\bar{\Psi}^{\lambda'}(r',t')
 ~e^{-iS[\Psi,\bar{\Psi},\Phi]}}{\displaystyle\int D\Psi
D\bar{\Psi}D\Phi
 ~e^{-iS[\Psi,\bar{\Psi},\Phi]}},
  \nonumber\\
\end{eqnarray}
where the action in real space is given by
\begin{widetext}
\begin{eqnarray}\label{eq:swithphi}
S[\Psi,\bar{\Psi},\Phi]=\int dt d^2r
\Big\{\sum_{\lambda\lambda'}\bar{\Psi}^{\lambda}(\mathbf{r},t)\big[W_{\lambda\lambda'}\sigma_3-
e\tilde{\phi}_i\tilde{\gamma}_i\delta_{\lambda\lambda'}\big]
\Psi^{\lambda'}(\mathbf{r},t)\Big\}
  \nonumber\\
+\int dt\int d^2r d^2r'
\Big\{\Phi^T(\mathbf{r},t)\frac{-e^2}{2}V_0^{-1}(\mathbf{r}-\mathbf{r'})\sigma_3\Phi(\mathbf{r'},t)\Big\},
\end{eqnarray}
\end{widetext}
in which the Pauli matrix $\sigma_3=\mathrm{diag}(1, -1)$ is defined
on the Keldysh space, and $V_0^{-1}$ is defined via the following
relation
\begin{eqnarray*}
\int d^2r_1
V_0(\mathbf{r}-\mathbf{r_1})V_0^{-1}(\mathbf{r_1}-\mathbf{r'})
=\delta(\mathbf{r}-\mathbf{r'}).
\end{eqnarray*}
The other notions appeared in Eq.~(\ref{eq:swithphi}) are
fermionic doublet $\Psi$, bosonic doublet $\Phi$,
and vertex matrices $\tilde{\gamma}_i$, they are defined as
\begin{eqnarray*}
&& \Psi^\lambda=\left( {\begin{array}{c}
   \psi^{\lambda}_1 \\
   \psi^{\lambda}_2
\end{array}} \right),\quad
    \Phi=\left( {\begin{array}{c}
    \tilde{\phi}_{1} \\
    \tilde{\phi}_{2}
\end{array}} \right),
      \\[2mm]
&&\tilde{\gamma}^{}_1=\left( {\begin{array}{*{100}c}
   1& 0 \\
   0& 0
\end{array}} \right), \quad
   \tilde{\gamma}^{}_2=\left( {\begin{array}{*{100}c}
   0& 0 \\
   0& -1
\end{array}} \right).\nonumber
\end{eqnarray*}

For calculation convenience,
one can introduce a partition function
for the coupling between the fermionic and bosonic doublets,
\begin{eqnarray*}
&& Z[\Phi]=\ave{\mathrm{T_C}~e^{-iS_R[\Phi,\Psi]}}_\Psi,
   \\
&& S_R[\Phi,\Psi]=\int dt\ d^2r\
\Big\{\sum_\lambda\bar{\Psi}^\lambda
 \big(-e\tilde{\phi}_i\tilde{\gamma}_i\big)\Psi^\lambda\Big\},
\end{eqnarray*}
where $\mathrm{T_C}$ stands for time ordering along the contour
$\mathrm{C}$ and $\ave{ \cdots }_\Psi$ means functional integration
over $\Psi$ field with the action
$$
S[\Psi]=\int dt d^2r
\Big\{\sum_{\lambda\lambda'}W_{\lambda\lambda'}\bar{\Psi}^{\lambda}(\mathbf{r},t)
\sigma_3 \Psi^{\lambda'}(\mathbf{r},t)\Big\}.
$$
Then the Green's function in Eq.~(\ref{eq:Green3})
can be formally expressed as a functional integration over the bosonic fields,
\begin{eqnarray}\label{eq:Green1}
\hat{G}_{\lambda\lambda'}(\mathbf{r},t;\mathbf{r}',t')&=&\mathbb{N}\int
D\Phi\,\hat{G}_{\lambda\lambda'}(\mathbf{r},t;\mathbf{r}',t'\mid
\Phi)\nonumber\\
&&\times \exp\big\{-iS_e[ \Phi]\big\},
\end{eqnarray}
where the normalization coefficient is denoted by $\mathbb{N}$
and the action $S_e[\Phi]$ is defined by
\begin{eqnarray}
 && S_e[\Phi]= i\ln Z[\Phi]+\int dt\ d^2r d^2r' \hspace{3mm}
 \nonumber\\
 &&\hspace{6mm}\times
  \big\{\Phi^T(\mathbf{r},t)
\frac{-e^2}{2}V_0^{-1}(\mathbf{r}-\mathbf{r'})\sigma_3
\Phi(\mathbf{r'},t)\big\},
\end{eqnarray}
and the kernel $\hat{G}(\mathbf{r},t;\mathbf{r}',t'\mid \Phi)$
is given by
\begin{eqnarray}\label{green3}
&&\hat{G}_{\lambda\lambda'}(\mathbf{r},t;\mathbf{r}',t'\mid
\Phi)\nonumber\\
&&=\frac{1}{Z[\Phi]}\langle \mathrm{T_C}\
\Psi^{\lambda}(\mathbf{r},t)
\bar{\Psi}^{\lambda'}(\mathbf{r'},t')e^{-iS_R[\Phi,\Psi]}\rangle_\Psi.
\end{eqnarray}
We can average the Green's function
$\hat{G}_{\lambda\lambda'}(\mathbf{r},t;\mathbf{r}',t')$ over
disorder as follows~\cite{Zala}
\begin{eqnarray}
\langle\hat{G}_{\lambda\lambda'}(\mathbf{r},t;\mathbf{r}',t')\rangle_{dis}&=&\mathbb{N}\int
D\Phi\,\langle\hat{G}_{\lambda\lambda'}(\mathbf{r},t;\mathbf{r}',t'\mid
\Phi)\rangle_{dis}\nonumber\\
&&\times \exp\big\{-i\langle S_e[ \Phi]\rangle_{dis}\big\},
\end{eqnarray}
where $\langle \cdots \rangle_{dis}$ refers to the average over
disorder. The random disorder potential $U(r)$ is assumed to be
characterized by a correlation function
\begin{eqnarray*}
\langle U(r)U(r')\rangle_{dis}=\frac{1}{2\pi\nu\tau}\delta(r-r'),
\end{eqnarray*}
where $\nu=m/\pi\hbar^2$ stands for the density of states. The
average of the Green's function over disorder introduces the elastic
scattering time $\tau$ which is relevant to the random disorder. We
neglect correlations between the mesoscopic fluctuations of $\langle
S_e[ \Phi]\rangle_{dis}$ and the fermionic operators in
Eq.~(\ref{green3}) so that the average of the Green's function
$\hat{G}_{\lambda\lambda'}(\mathbf{r},t;\mathbf{r}',t'\mid \Phi)$
can be separated from the bosonic action $S_e[ \Phi]$. This
approximation is valid since the mesoscopic fluctuation is smaller
than average quantities.

After averaging over disorder, we rotate the Keldysh bases
$\hat{G}\rightarrow L\sigma_3 \hat{G} L^{\dag}$ through a unitary
matrix $L$
\begin{eqnarray*}
L=\frac{1}{\sqrt{2}}\left( {\begin{array}{*{100}c}
   1& -1 \\
   1& 1
\end{array}} \right),
\end{eqnarray*}
so that the Green's function takes the following shape,
\begin{eqnarray}\label{eq:Green2}
&&\hat{G}(\mathbf{r},t;\mathbf{r}',t'\mid
\Phi)\nonumber\\
&&=\left( {\begin{array}{*{100}c}
   G^R(\mathbf{r},t;\mathbf{r}',t'\mid
\Phi)& G^K(\mathbf{r},t;\mathbf{r}',t'\mid
\Phi) \\
   G^Z(\mathbf{r},t;\mathbf{r}',t'\mid
\Phi)& G^A(\mathbf{r},t;\mathbf{r}',t'\mid \Phi)
\end{array}} \right).
\end{eqnarray}
Note that the Green's function
$\hat{G}(\mathbf{r},t;\mathbf{r}',t'\mid \Phi)$ is
a $2\times 2$ matrix defined in the Keldysh space,
of which the matrix entities are again
$2\times 2$ matrices defined in spin space.

The bosonic fields after rotation take the following two components
$\phi_1=\displaystyle\frac{e}{2}(\tilde{\phi}_1+\tilde{\phi}_2)$ and
$\phi_2=\displaystyle\frac{e}{2} (\tilde{\phi}_1-\tilde{\phi}_2)$
that reside on the upper and lower branches of the contour
$\mathrm{C}$, respectively. Then the corresponding vertex matrices
turn to
$\gamma_{1(2)}=L(\tilde{\gamma}_1\pm\tilde{\gamma}_2)\sigma_3L^{\dag}$,
namely,
\begin{eqnarray*}
\gamma_{1}=\left( {\begin{array}{*{100}c}
   1& 0 \\
   0& 1
\end{array}} \right),\quad
\gamma_{2}=\left( {\begin{array}{*{100}c}
   0& 1 \\
   1& 0
\end{array}} \right),
\end{eqnarray*}
and the interaction term
$\bar{\Psi}^\lambda(-e\tilde{\phi}_i\tilde{\gamma}_i)\Psi^\lambda$
becomes
$-\bar{\Psi}^\lambda\left( {\begin{array}{cc}
   \phi_1& \phi_2 \\
   \phi_2& \phi_1
\end{array}} \right)\Psi^\lambda$.
As $\gamma_1$ and $\gamma_2$ constitute a representation of $Z_2$
group, the interaction can be regarded as the coupling between Fermi
field and $Z_2$ Bose field.

These Bose fields define the following propagators:
\begin{eqnarray}
&&D^K(\mathbf{r_1},\mathbf{r_2};t_1,t_2)=-2i\langle\phi_1(\mathbf{r_1},t_1)\phi_1(\mathbf{r_2},t_2)\rangle,\nonumber\\
&&D^R(\mathbf{r_1},\mathbf{r_2};t_1,t_2)=-2i\langle\phi_1(\mathbf{r_1},t_1)\phi_2(\mathbf{r_2},t_2)\rangle ,\nonumber\\
&&D^A(\mathbf{r_1},\mathbf{r_2};t_1,t_2)=-2i\langle\phi_2(\mathbf{r_1},t_1)\phi_1(\mathbf{r_2},t_2)\rangle ,\nonumber\\
&&\langle\phi_2(\mathbf{r_1},t_1)\phi_2(\mathbf{r_2},t_2)\rangle=0.
\end{eqnarray}
One can show that those propagators
obey the Dyson equations in the saddle point approximation,
namely,
\begin{eqnarray}\label{polar2}
\hat{D}(x_1,x_2)=\hat{D}_0 + \!\int\!\! dx_3dx_4
\hat{D}_0(x_1,x_3)\hat{\Pi}(x_3,x_4)\hat{D}(x_3,x_2),
 \nonumber\\
\end{eqnarray}
with notations $x \equiv(\mathbf{r},t)$,
$D_0^R(q)=D_0^A(q)=-2\pi e^2/|\mathbf{q}|$, and
\begin{widetext}
\begin{eqnarray*}
\hat{D}=\left(%
\begin{array}{cc}
  D^R & D^K \\
  0   & D^A \\
\end{array}%
\right),\quad
\hat{D}_0=\left( {\begin{array}{*{100}c}
   D_0^R& 0 \\
   0& D_0^A
\end{array}} \right),\quad
   \hat{\Pi}=\left( {\begin{array}{*{100}c}
   \Pi^R& \Pi^K \\
   0& \Pi^A
\end{array}} \right), \nonumber
\end{eqnarray*}
\begin{eqnarray}\label{polar}
&&\Pi^R(x_1,x_2)=\Pi^A(x_2,x_1)=\frac{\delta\
\mathrm{Tr_s}G^K(\mathbf{r}_1;t_1,t_1\mid \Phi)}{-2i\ \delta
\phi_1(\mathbf{r}_2,t_2)},\nonumber\\
&&\Pi^K(x_1,x_2)=\frac{\delta\
\mathrm{Tr_s}\big[G^K(\mathbf{r}_1;t_1,t_1\mid
\Phi)+G^Z(\mathbf{r}_1;t_1,t_1\mid \Phi)\big]}{-2i\ \delta
\phi_2(\mathbf{r}_2,t_2)},\nonumber\\
\end{eqnarray}
\end{widetext}
where $\mathrm{Tr_s}$ stands for the trace in the spin space.

\section{The kinetic equation}\label{sec:kinetic}

In previous section, the electron-electron interaction has been
decoupled with the help of auxiliary bosonic fields $\phi_{1(2)}$.
This means that the influence of the interaction can be described by
a $Z_2$ Bose field,
\begin{eqnarray}
\hat{\varphi}(\mathbf{r},t)=\left( {\begin{array}{*{100}c}
   \phi_1(\mathbf{r},t)& \phi_2(\mathbf{r},t) \\
   \phi_2(\mathbf{r},t)& \phi_1(\mathbf{r},t)
\end{array}} \right).\nonumber
\end{eqnarray}
Now we are able to apply the quasiclassical Green's
function~\cite{Eilenberger,Rammer,Schwab} approach to study the spin
dynamics. We derive the Eilenberger equation from the right-hand and
left-hand Dyson equations obeyed by the Green's function
$\hat{G}(\mathbf{r},t;\mathbf{r}',t'\mid \Phi)$ in
Eq.~(\ref{eq:Green2}):
\begin{eqnarray}\label{eq:quasiGreen0}
\tilde{\partial}_t \hat{g}+\mathbf{v_F}\cdot \vec{\nabla}
\hat{g}+i\big[\mathbf{b}\cdot\vec{\sigma},\hat{g}\big]=
\frac{\hat{g}~\langle~\hat{g}~\rangle_{\mathbf{n}}-\langle~\hat{g}~\rangle_{\mathbf{n}}~\hat{g}}{2\
\tau},
\end{eqnarray}
where $\mathbf{v}_F$ denotes the Fermi velocity, $\tau$ is the
elastic scattering time arising from the adoption of the standard
self-consistent Born approximation; $\langle \cdots\rangle_{\mathbf
n}$ means taking average over the direction of the electron momentum
$\mathbf{n}=\mathbf{p}/|\mathbf{p}|\equiv(\cos\theta,~\sin\theta)$,
and the covariant derivative is defined by
\begin{eqnarray}
\tilde{\partial}_t \hat{g}=\partial_{t_1}\hat{g}+\partial
_{t_2}\hat{g}+i\hat{\varphi}(\mathbf{r},t_1)\hat{g}-i\hat{g}
\hat{\varphi}(\mathbf{r},t_2).
\end{eqnarray}
The quasiclassical Green's function in Keldysh and spin spaces,
\begin{eqnarray}\label{eq:quasiGreen1}
\hat{g}=\left( {\begin{array}{*{100}c} g^R &g^K\\
g^Z &g^A
\end{array}}\right),
\end{eqnarray}
can be derived by integrating the Fourier transform
of the Green's function in Eq.~(\ref{eq:Green2}) over energy variables,
\ie,
\begin{eqnarray}\label{eq:quasiGreen3}
\hat{g}(t_1,t_2;\mathbf{n},\mathbf{r})=\frac{i}{\pi}\int d\xi
\hat{G}(t_1,t_2;\mathbf{p},\mathbf{r}),\hspace{16mm}
   \nonumber\\
\hat{G}(t_1,t_2;\mathbf{p},\mathbf{r})=\int d^2 r'
e^{i\mathbf{p}\cdot\mathbf{r}'}
\hat{G}(\mathbf{r}_1,t_1;\mathbf{r}_2,t_2\mid
\Phi),
\end{eqnarray}
where $\xi=\mathbf{p}^2/2m-\mu$,
$\mathbf{r}'=\mathbf{r}_1-\mathbf{r}_2$,
$\mathbf{r}=(\mathbf{r}_1+\mathbf{r}_2)/2$.
The electron polarization operators can be obtained in terms of
Eq.~(\ref{polar}) and Eq.~(\ref{eq:quasiGreen3}), \ie,
\begin{eqnarray}\label{polar1}
&&\hspace{4mm}\Pi^R(x_1,x_2)=\Pi^A(x_2,x_1)\nonumber\\
&&=\nu \int \frac{d\theta}{2\pi}\big[\ \delta(x_1-x_2)+\frac{\pi\
\delta\
\mathrm{Tr_s}g^K(t_1,t_1;\mathbf{n},\mathbf{r}_1)}{2\ \delta\phi_1(\mathbf{r}_2,t_2)}\big],\nonumber\\
&&\hspace{4mm}\Pi^K(x_1,x_2)\nonumber\\
&&=\pi\nu \int \frac{d\theta}{2\pi}\frac{\delta\
\mathrm{Tr_s}\big[g^K(t_1,t_1;\mathbf{n},\mathbf{r}_1)+g^Z(t_1,t_1;\mathbf{n},\mathbf{r}_1)\big]}{2\
\delta
\phi_2(\mathbf{r}_2,t_2)}.\nonumber\\
\end{eqnarray}

Since physical observables are determined by the Keldysh component
of the quasiclassical Green's function, namely
$\ave{g^K(t_1,t_2;\mathbf{n},\mathbf{r})}_{\Phi}$  (here the
subscript $\Phi$ refers that the functional average~\cite{onAverage}
is taken over the field $\Phi$), we need to solve this component
from the Eilenberger equation. Decomposing the Green's function in
charge and spin components
$\ave{g^K(t_1,t_2;\mathbf{n},\mathbf{r})}_{\Phi}
 = g^K_0 + \mathbf{g}^K \cdot \vec{\sigma}$,
one can obtain the charge and spin densities, respectively,
\begin{eqnarray}
\rho(\mathbf{r},t)&=&-\frac{1}{4}e\nu\int d\epsilon
 \ave{g^K_0(t,\epsilon;\mathbf{n},\mathbf{r})}_{\mathbf{n}},
  \nonumber\\
\mathbf{S}(\mathbf{r},t)&=&-\frac{1}{4}\nu\int d\epsilon \langle
  \mathbf{g}^{K}(t,\epsilon;\mathbf{n},\mathbf{r})\rangle_{\mathbf{n}}.
\end{eqnarray}

Now we turn to the kinetic equations for the two independent
components $g^K$ and $g^Z$.
For $\ave{ g^Z }_\Phi = 0$ in all orders of the perturbation theory,
we have
$$
g^K=\ave{ g^K }_\Phi + \delta g^K,\quad
g^Z=\delta g^Z,
$$
where the fluctuation parts $\delta g$ imply the effects contributed
by the auxiliary bosonic fields. One can obtain from
Eq.~(\ref{eq:quasiGreen0}) that $\delta g^Z$ obeys the following
equation,
\begin{eqnarray}\label{eq1}
&&(\tilde{\partial}_t +\mathbf{v}^{}_F\cdot \vec{\nabla})\delta
g^Z+i\big[\mathbf{b}\cdot\vec{\sigma},\delta
g^Z\big]-\frac{1}{\tau}\big[\delta g^Z-\langle\delta g^Z\rangle_\mathbf{n}\big]\nonumber \\
&&=-2i\phi_2(\mathbf{r},t_1)\delta(t_1-t_2)\mathrm{I_s},
\end{eqnarray}
where $\mathrm{I_s}$ denotes the unit matrix in spin space.
When deriving the above equation,
we have used the conditions
$g^R=\delta(t_1-t_2)\mathrm{I_s}-g^K\delta
g^Z/2$ and $g^A=-\delta(t_1-t_2)\mathrm{I_s}+\delta g^Z g^K/2$.
Equation (\ref{eq1}) gives rise to
\begin{eqnarray}\label{dgz}
&&\delta g^Z(t_1,t_2;\mathbf{n},\mathbf{r})
 = 2i\delta(t_1-t_2)\int d\mathbf{r}_1dt_3\int\frac{d\theta'}{2\pi}
    \nonumber\\
 &&\hspace{25mm}\times\phi_2(\mathbf{r}_1,t_3)
  \Gamma_\rho(t_3-t_1,\mathbf{n}',\mathbf{n};
\mathbf{r}_1,\mathbf{r}),
   \nonumber\\
&&\Gamma_\rho(t,\mathbf{n}',\mathbf{n}; \mathbf{r}_1,\mathbf{r}_2)
 =\int\frac{d\omega d^2\mathbf{q}}{(2\pi)^3}e^{i\mathbf{q}\cdot
(\mathbf{r}_1-\mathbf{r}_2)-i\omega t}\nonumber\\
&&\hspace{30mm}\times\Gamma_\rho(\mathbf{n}',\mathbf{n};\omega,\mathbf{q}),
\end{eqnarray}
where the diffusion propagator $\Gamma_\rho$ is defined by
\begin{eqnarray}\label{gm1}
&&(-i\omega+iv_F\mathbf{n}\cdot
\mathbf{q})\Gamma_\rho(\mathbf{n},\mathbf{n}';
\omega,\mathbf{q})+\frac{1}{\tau}\big[\Gamma_\rho(\mathbf{n},\mathbf{n}';
\omega,\mathbf{q})\nonumber\\
&&-\langle\Gamma_\rho(\mathbf{n},\mathbf{n}';
\omega,\mathbf{q})\rangle_\mathbf{n}\big]=2\pi\delta(\mathbf{n}-\mathbf{n}').
\end{eqnarray}
After obtaining the explicit form of $\delta g^Z$,
we can further solve the $\delta g^K$ from the
following relation
\begin{eqnarray}\label{eq:quasiGreen2}
&&(\tilde{\partial}_t +\mathbf{v_F}\cdot \vec{\nabla})\delta
g^K+i\big[\mathbf{b}\cdot\vec{\sigma},\delta
g^K\big]+\frac{1}{\tau}\big[\delta g^K-\langle\delta g^K\rangle_\mathbf{n}\big]\nonumber \\
&=&2i\phi_2(\mathbf{r},t_1)\delta(t_1-t_2)I_s-i\big[\phi_1(\mathbf{r},t_1)
-\phi_1(\mathbf{r},t_2)\big]\langle g^K\rangle_\Phi\nonumber\\
&&+\frac{1}{4\tau}\Big[\ \langle g^K\rangle_\Phi\langle\ \delta
g^Z\langle g^K\rangle_\Phi\ \rangle_\mathbf{n}-\langle\ \langle
g^K\rangle_\Phi\ \rangle_\mathbf{n}\delta
g^Z\langle g^K\rangle_\Phi\nonumber\\
&&-\langle g^K\rangle_\Phi\delta g^Z\langle\ \langle
g^K\rangle_\Phi\ \rangle_\mathbf{n}+\langle\ \langle
g^K\rangle_\Phi\delta g^Z\ \rangle_\mathbf{n}\langle
g^K\rangle_\Phi\ \Big].
\end{eqnarray}
We take only the zeroth and first angular harmonics into account in
the Keldysh component assumed spatial smoothness,
\begin{eqnarray}\label{fj}
&&\ave{g^K(t_1,t_2;\mathbf{n},\mathbf{r})}_\Phi
  \approx\ave{g^K(t_1,t_2;\mathbf{n}',\mathbf{r})}_{\Phi, \mathbf{n}'}
   \nonumber\\
&& + 2\mathbf{n}\cdot
   \ave{\mathbf{n}'g^K(t_1,t_2;\mathbf{n}',\mathbf{r})}_{\Phi, \mathbf{n}'}.
\end{eqnarray}

Decomposing the fluctuating term in charge and spin components
$\delta g^K=\delta g^K_0+\delta\mathbf{g}^{K}\cdot \vec{\sigma}$,
one can easily obtain the explicit expression of the $\delta g^K_0$
which is given in Eq.~(\ref{dgk1}). The fluctuation part
$\delta\mathbf{g}^K$ related to the spin components fulfils the
following equation,
\begin{widetext}
\begin{eqnarray}\label{gk3}
(\tilde{\partial}_t +\mathbf{v_F}\cdot \vec{\nabla})\delta
\mathbf{g}^K-2\mathbf{b}\times\delta
\mathbf{g}^K+\frac{1}{\tau}\big[\delta \mathbf{g}^K-\langle\delta
\mathbf{g}^K\rangle_\mathbf{n}\big]
=-i\big[\phi_1(\mathbf{r},t_1)-\phi_1(\mathbf{r},t_2)\big]
\Big(\langle\mathbf{g}^K(t_1,t_2;\mathbf{n},\mathbf{r})\rangle_{\Phi, \mathbf{n}}
      \hspace{9mm}\nonumber\\
+2\mathbf{n}\cdot\langle\mathbf{n}'\mathbf{g}^K(t_1,t_2;\mathbf{n}',\mathbf{r})\rangle_{\Phi,
\mathbf{n}'}\Big) +\frac{1}{2\tau}\Big\{\langle g^K_0\rangle_{\Phi,
\mathbf{n}}\big(\langle\delta g^Z\rangle_{\mathbf{n}}-\delta
g^Z\big)\langle\mathbf{g}^K\rangle_{\Phi,
\mathbf{n}}+\langle\mathbf{g}^K\rangle_{\Phi,
\mathbf{n}}\big(\langle\delta g^Z\rangle_\mathbf{n}-\delta
g^Z\big)\langle g^K_0\rangle_{\Phi, \mathbf{n}}\Big\}.
\end{eqnarray}
\end{widetext}
If denoting
\begin{eqnarray}\label{juzhen1}
Q=\left( {\begin{array} {*{100}c}
 \delta g^{K}_x\\[1.5mm]
 \delta g^{K}_y\\[1.5mm]
 \delta g^{K}_z
 \end{array}}\right),
     \quad
L_k=\left( {\begin{array} {*{100}c}
  g^K_x\\
  g^K_y\\
  g^K_z\end{array}}\right),\nonumber
\end{eqnarray}
we can write Eq.~(\ref{gk3}) in the following matrix equation,
\begin{eqnarray}\label{juzhen2}
(\tilde{\partial}_t +\mathbf{v_F}\cdot
\vec{\nabla})Q-2\zeta Q+\frac{1}{\tau}\big(Q-\langle Q\rangle_{\mathbf{n}}\big)
   \hspace{18mm}\nonumber \\
= -i\big[\phi_1(\mathbf{r},t_1)-\phi_1(\mathbf{r},t_2)\big]
\big(\langle L_k\rangle_{\Phi, \mathbf{n}}
+2\mathbf{n}\cdot\langle\mathbf{n}'L_k\rangle_{{\Phi, \mathbf{n}'}}\big)
   \nonumber\\
+\frac{1}{2\tau}\Big\{\langle g^K_0\rangle_{\Phi,
\mathbf{n}}\big(\langle\delta g^Z\rangle_\mathbf{n}-\delta
g^Z\big)\langle L_k\rangle_{\Phi, \mathbf{n}}
  \hspace{18mm}\nonumber\\
+\langle L_k\rangle_{\Phi, \mathbf{n}}\big(\langle\delta
g^Z\rangle_\mathbf{n}-\delta g^Z\big)\langle g^K_0\rangle_{\Phi,
\mathbf{n}}\Big\},\hspace{21mm}
\end{eqnarray}
where the matrix $\zeta$ is given by
\begin{eqnarray}
 \zeta=\left( {\begin{array}
{*{100}c}
 0& 0 & -\alpha p_F\cos\theta\\
0& 0 & -\alpha p_F\sin\theta\\
 \alpha p_F\cos\theta& \alpha p_F\sin\theta &
 0\end{array}}\right).\nonumber
\end{eqnarray}
Then equation (\ref{juzhen2}) can be solved by utilizing  the following
expression
\begin{eqnarray}
&&(-i\omega+iv_F\mathbf{n}\cdot
\mathbf{q}-2\zeta)\Gamma_s(\mathbf{n},\mathbf{n}';
\omega,\mathbf{q})+\frac{1}{\tau}\big[\Gamma_s(\mathbf{n},\mathbf{n}';
\omega,\mathbf{q})\nonumber\\
&&-\langle\Gamma_s(\mathbf{n},\mathbf{n}';
\omega,\mathbf{q})\rangle_\mathbf{n}\big]=2\pi\delta(\mathbf{n}-\mathbf{n}').
\end{eqnarray}
We give the explicit expression of $\delta \mathbf{g^{K}}$ in
Eq.~(\ref{dgk2}) in the appendix~\ref{sec:explicit}.

Since the concrete forms of $\delta g^Z$ and $\delta g^K$ have
been obtained, we can write down the kinetic equation satisfied by
the Keldysh function through averaging the $K$ component of
Eq.~(\ref{eq:quasiGreen0}) over the auxiliary bosonic fields,
\begin{eqnarray}\label{kinetic1}
&&(\tilde{\partial}_t +\mathbf{v_F}\cdot \vec{\nabla})
\langle g^K\rangle_\Phi+i\big[\mathbf{b}\cdot\vec{\sigma},\langle g^K\rangle_\Phi\big]\nonumber\\
&&=C_{el}\big\{\langle g^K\rangle_\Phi\big\}+C_{in}\big\{\langle
g^K\rangle_\Phi\big\},
\end{eqnarray}
where the inelastic collision integral reads
\begin{eqnarray}\label{cin}
&&C_{in}\{\langle g^K\rangle_\Phi\}(t_1,t_2;\mathbf{n},\mathbf{r})\nonumber\\
&&=-i\langle\
\big[\phi_1(\mathbf{r},t_1)-\phi_1(\mathbf{r},t_2)\big]\delta g^K\
\rangle_\Phi,
\end{eqnarray}
and the elastic collision integral is given by
\begin{widetext}
\begin{eqnarray}\label{ke1}
&&C_{el}\big\{\langle
g^K\rangle_\Phi\big\}(t_1,t_2;\mathbf{n},\mathbf{r})=\frac{1}{\tau}\big[\langle\
\langle g^K(t_1,t_2;\mathbf{n},r)\rangle_\Phi\ \rangle_\mathbf{n}-
\langle g^K(t_1,t_2;\mathbf{n},\mathbf{r})\rangle_\Phi\big]\nonumber\\
&&\hspace{3mm}+\int dt_3\frac{d\theta_1}{2\pi}\big[\langle
g^K(t_1,t_3;\mathbf{n}_1,\mathbf{r})\rangle_\Phi
\Lambda^A(t_3,t_2;\mathbf{n}_1,\mathbf{n};\mathbf{r})-\langle
g^K(t_1,t_3;\mathbf{n},\mathbf{r})\rangle_\Phi
\Lambda^A(t_3,t_2;\mathbf{n},\mathbf{n}_1;\mathbf{r})\big]\nonumber\\
&&\hspace{3mm}+\int
dt_3\frac{d\theta_1}{2\pi}\big[\Lambda^R(t_1,t_3;\mathbf{n},\mathbf{n}_1;\mathbf{r})
\langle
g^K(t_3,t_2;\mathbf{n}_1,\mathbf{r})\rangle_\Phi-\Lambda^R(t_1,t_3;\mathbf{n}_1,\mathbf{n};\mathbf{r})
\langle g^K(t_3,t_2;\mathbf{n},\mathbf{r})\rangle_\Phi \big],
\end{eqnarray}
where
\begin{eqnarray}
\Lambda^A(t_1,t_2;\mathbf{n},\mathbf{n}_1;\mathbf{r})
 =\frac{1}{4\tau}\int dt_3\langle\
 \big[\delta g^Z(t_1,t_3;\mathbf{n}_1,\mathbf{r})
  -\delta g^Z(t_1,t_3;\mathbf{n},\mathbf{r})\big]\delta
g^K(t_3,t_2;\mathbf{n},\mathbf{r})\ \rangle_{\Phi},
       \nonumber\\
\Lambda^R(t_1,t_2;\mathbf{n},\mathbf{n}_1;\mathbf{r})=\frac{1}{4\tau}
\int dt_3\langle\delta g^K(t_1,t_3;\mathbf{n},\mathbf{r})
\big[\delta g^Z(t_3,t_2;\mathbf{n}_1,\mathbf{r})-\delta
g^Z(t_3,t_2;\mathbf{n},\mathbf{r})\big]\rangle_{\Phi}.
\end{eqnarray}
\end{widetext}
Substituting the explicit forms of $\delta g^Z$ and $\delta g^K$
given in Eq.~(\ref{dgz}), (\ref{dgk1}) and (\ref{dgk2}) into
Eq.~(\ref{kinetic1}), one get the kinetic equation which can be used
to study the influence of electron-electron interaction on the spin
dynamics of 2DEGs with Rashba spin-orbit coupling. After some
tedious calculation, we obtain the explicit expressions of the
inelastic and elastic collision integrals, respectively, which are
given in Eq.~(\ref{c1}) and (\ref{c2}) in the
appendix~\ref{sec:explicit}.

\section{spin dynamics}\label{sec:dynamics}

After taking average over the direction of the momentum $\mathbf{n}$,
one can see from Eq.~(\ref{cin}-\ref{ke1}) that the elastic collision integral
vanishes,
\begin{eqnarray}\label{cin3}
\int\frac{d\theta}{2\pi}C_{el}\big\{\langle
g^K\rangle_\Phi\big\}(t,\epsilon;\mathbf{n},\mathbf{r})=0,
\end{eqnarray}
but the average of the inelastic collision integral over the
direction does not vanish. This means that the elastic collision
integral preserves the number of electrons on a given energy shell
defined by Eq.~(\ref{bh}), while the inelastic collision integral
does not preserve it. When $t_1=t_2$, equation~(\ref{cin2}) gives
rise to
\begin{eqnarray*}\label{cin0}
\int\! d\epsilon C_{in}\big\{\langle
   g^K\rangle_\Phi\big\}(t_1,\epsilon;\mathbf{n},\mathbf{r})
  &=&C_{in}\big\{\langle
  g^K\rangle_\Phi\big\}(t_1,t_1;\mathbf{n},\mathbf{r}).
\end{eqnarray*}
One can see from Eq.~(\ref{cin}) that the right-hand side is always
zero. Thus we obtain
\begin{eqnarray}\label{cin0}
\int d\epsilon C_{in}\big\{\langle
g^K\rangle_\Phi\big\}(t,\epsilon;\mathbf{n},\mathbf{r})=0.
\end{eqnarray}
This implies that
not only the total number of electrons is conserved,
but also the number of electrons moving along a concrete direction
$\mathbf{n}$ is conserved.

Decomposing the Green's function in charge and spin components in
the approximation of Eq.~(\ref{fj}), separating the zeroth and first
angular harmonics and utilizing Eq.~(\ref{cin3}) and
Eq.~(\ref{cin0}), we obtain from Eq.~(\ref{kinetic1}) that
\begin{eqnarray}\label{kinetic2}
&&v_F\mathbf{n}\cdot\vec{\nabla}\langle
g^K(t,\epsilon;\mathbf{n},\mathbf{r})\rangle_{\Phi, \mathbf{n}}+
i\big[\mathbf{b}\cdot\vec{\sigma},\nonumber\\
&&\langle~g^K(t,\epsilon;\mathbf{n},\mathbf{r})~\rangle_{\Phi,
\mathbf{n}}\big]= C_{el}\big\{\langle
g^K\rangle_\Phi\big\}(t,\epsilon;\mathbf{n},\mathbf{r}),
\end{eqnarray}
\begin{eqnarray}\label{kinetic3}
&&\partial_t\langle
g^K(t,\epsilon;\mathbf{n},\mathbf{r})\rangle_{\Phi,
\mathbf{n}}+i\big[\mathbf{b}\cdot\vec{\sigma},
2\mathbf{n}\cdot\langle\mathbf{n}'g^K(t,\epsilon;\mathbf{n}',\mathbf{r})\rangle_{\Phi, \mathbf{n}'}\big] \nonumber\\
&&+v_F\mathbf{n}\cdot\vec{\nabla}\big(2\mathbf{n}\cdot\langle\
\mathbf{n}'g^K(t,\epsilon;\mathbf{n}',\mathbf{r})\ \rangle_{\Phi,
\mathbf{n}'}\big)=0.
\end{eqnarray}
There is no contribution of the inelastic collision integral to the
spin dynamics due to the condition Eq.~(\ref{cin0}). Solving
Eq.~(\ref{kinetic2}) and substituting
$\langle\mathbf{n}'g^K(t,\epsilon;\mathbf{n}',\mathbf{r})\rangle_{\Phi,
\mathbf{n}'}$ into Eq.~(\ref{kinetic3}), we can obtain the spin and
charge diffusion equations. The diffusion equation for the charge
density reads,
\begin{eqnarray}
&&\partial_t\rho-C_D\partial_\mathbf{X}^2\rho=0,
\end{eqnarray}
where $\mathbf{\partial_X}=(\partial_x,\partial_y)$ and
$C_D=v_F^2\tau/2$.
We introduce the distribution function $f$ which
reduces to the Fermi distribution in equilibrium,
\begin{eqnarray}
f=f_0+\vec{\sigma} \cdot
\mathbf{f}_\mathbf{k}=\frac{1}{2}(1-\frac{1}{2}g^K).
\end{eqnarray}
In the time $\tau$, the charge density becomes isotropic but the
spin relaxation process does not start yet, hence~\cite{Dyakonov2}
\begin{eqnarray}
&&g^K_0(\epsilon)=2(1-2f_0(\epsilon)),\nonumber\\
&&\mathbf{g}^K
(t,\epsilon;\mathbf{r})=-4\mathbf{f}_\mathbf{k}(t,\epsilon;\mathbf{r}),
\end{eqnarray}
where
\begin{eqnarray}
&&f_0(\epsilon)=\big(f_+(\epsilon)+f_-(\epsilon)\big)/2,\nonumber\\
&&\mathbf{f}_\mathbf{k}
(t,\epsilon;\mathbf{r})=\big(f_+(\epsilon)-f_-(\epsilon)\big)\mathbf{s}(t,\mathbf{r}),\nonumber\\
&&f_{\pm}(\epsilon)=[\exp(\frac{\epsilon\mp\Delta
\mu/2}{k_BT})+1]^{-1},
\end{eqnarray}
where $\mathbf{s} = (s_x, s_y, s_z)$ denotes the unit vector along
the spin,  $f_{\pm}(\epsilon)$ represent the distribution functions
projected along the direction parallel or antiparallel to the unit
vector $\mathbf{s}$ (all the energies are counted from the Fermi
energy) and $\Delta \mu=(\mu_+-\mu_-)$ refers to the difference
between the chemical potentials $\mu_{\pm}$ of the electron spin
subsystems. The diffusion equations for the spin components are
given by
\begin{eqnarray}\label{diff1}
&&\partial_tS_x-C_D\partial_\mathbf{X}^2S_x-2C_E\partial_xS_z+\frac{1}{\tau_s'}S_x\nonumber\\
&&~~~=\frac{1}{\tau_{xx}^e}S_x+F_x(S_x, S_y, S_z),\nonumber\\
&&\partial_tS_y-C_D\partial_\mathbf{X}^2S_y-2C_E\partial_yS_z+\frac{1}{\tau_s'}S_y\nonumber\\
&&~~~=\frac{1}{\tau_{yy}^e}S_y+F_y(S_x, S_y, S_z),\nonumber\\
&&\partial_tS_z-C_D\partial_\mathbf{X}^2S_z+2C_E\partial_xS_x+2C_E\partial_yS_y+\frac{2}{\tau_s'}S_z\nonumber\\
&&~~~=\frac{1}{\tau_{zz}^e}S_z+F_z(S_x, S_y, S_z),
\end{eqnarray}
where $C_E=\alpha v_Fp_F\tau$, $\tau_s'=1/[2(\alpha p_F)^2\tau]$ and
$F_\ell(S_x, S_y, S_z)$ is a quadratic form of $(S_x, S_y, S_z)$
lacking of the $S^2_\ell$ $(\ell=x, y, z)$ term. The characteristic
times $\tau_{\ell\ell}^e$ describe the effect of the
electron-electron interaction on the spin relaxation, and their
explicit expressions are given by
\begin{eqnarray}\label{time1}
\frac{1}{\tau^e_{xx}}&=&\frac{2(\alpha p_F)^2\tau}{M}\Big\{\int
d\epsilon\int\frac{d\omega}{2\pi}\big[\big(f_+(\epsilon-\omega)
-f_-(\epsilon-\omega)\big)\nonumber\\
&\times& \mathrm{Im}(R^{xx}_2)g^K_0(\epsilon)+\big(f_+(\epsilon)
-f_-(\epsilon)\big) R^{xx}_1g^K_0(\epsilon-\omega)\big]\Big\},\nonumber\\
\frac{1}{\tau^e_{yy}}&=&\frac{2(\alpha p_F)^2\tau}{M}\Big\{\int
d\epsilon\int\frac{d\omega}{2\pi}\big[\big(f_+(\epsilon-\omega)
-f_-(\epsilon-\omega)\big)\nonumber\\
&\times& \mathrm{Im}(R^{yy}_2)g^K_0(\epsilon)+\big(f_+(\epsilon)
-f_-(\epsilon)\big) R^{yy}_1g^K_0(\epsilon-\omega)\big]\Big\},\nonumber\\
\frac{1}{\tau^e_{zz}}&=&\frac{1}{\tau^e_{xx}}+\frac{1}{\tau^e_{yy}}\nonumber\\
&=&\frac{2(\alpha p_F)^2\tau}{M}\Big\{\int
d\epsilon\int\frac{d\omega}{2\pi}\big[\big(f_+(\epsilon-\omega)
-f_-(\epsilon-\omega)\big)\nonumber\\
&\times& \big(\mathrm{Im}(R^{xx}_2)+\mathrm{Im}(R^{yy}_2)\big)g^K_0(\epsilon)\nonumber\\
&&+\big(f_+(\epsilon)
-f_-(\epsilon)\big) \big(R^{xx}_1+R^{yy}_1\big)g^K_0(\epsilon-\omega)\big]\Big\},\nonumber\\
\end{eqnarray}
where $M=\displaystyle\int d\epsilon \big(f_+(\epsilon)
-f_-(\epsilon)\big)$.
In order to obtain the concrete expressions of
the characteristic times $\tau_{\ell\ell}^e$,
we firstly take the
energy integration in Eq.~(\ref{time1}).
Since the spin splitting is
small, i.e.,
\begin{eqnarray*}
\mid \mu_+-\mu_- \mid \ll \mid \mu_+ \mid, \; \mid \mu_- \mid,
\end{eqnarray*}
the energy integration can be taken as follows,
\begin{eqnarray}
&&\frac{1}{M}\int_{-\infty}^\infty
d\epsilon\big(f_+(\epsilon-\omega) -f_-(\epsilon-\omega)\big)
\big(f_+(\epsilon)
+f_-(\epsilon)\big)\nonumber\\
&&\approx\frac{\displaystyle 2\int_{-\infty}^\infty
d\epsilon\frac{\partial
f_0(\epsilon-\omega)}{\partial\epsilon}f_0(\epsilon)}{\displaystyle\int_{-\infty}^\infty
d\epsilon\frac{\partial f_0(\epsilon)}{\partial\epsilon}}\nonumber\\
&&=1-\frac{\partial}{\partial \omega}(\omega
\coth\frac{\omega}{2k_BT}).
\end{eqnarray}
After the energy integration, the characteristic times
$\tau_{\ell\ell}^e$ have the forms
\begin{eqnarray}
&&\frac{1}{\tau^e_{xx}}=\frac{1}{\tau^e_{yy}}=8(\alpha
p_F)^2\tau\int_0^\infty\frac{d\omega}{2\pi}\big[\frac{\partial}{\partial
\omega}(\omega \coth\frac{\omega}{2k_BT})\big]\nonumber\\
&&\hspace{2.2cm}\times\big[\mathrm{Im}(R^{xx}_2)-R^{xx}_1\big],
\end{eqnarray}
the detail of the calculations of the kernels $R_1^{\imath\jmath}$
and $\mathrm{Im}R_2^{\imath\jmath}$ are given in appendix B. Now we
discuss the influence of the electron-electron interaction on the
spin-relaxation time in the ballistic regime $T\tau \gg 1$.

We can obtain the characteristic time $\tau_{xx}^e$ in the ballistic
regime utilizing the kernels $R_1^{\imath\jmath}$ and
$\mathrm{Im}R_2^{\imath\jmath}$ in Eq.~(\ref{ballistic})
\begin{eqnarray}\label{time3}
&&\frac{1}{\tau^e_{xx}}(T\tau \gg 1)=8(\alpha
p_F)^2\tau\int_0^\infty\frac{d\omega}{2\pi}\big[\frac{\partial}{\partial
\omega}(\omega \coth\frac{\omega}{2k_BT})\big]\nonumber\\
&&\hspace{2.2cm}\times\big[\frac{-1}{4\pi\nu
v_F^2}(\frac{3\pi}{2}+\tan^{-1}\omega\tau-\frac{2\omega\tau}{1+\omega^2\tau^2})\big]\nonumber\\
&&\approx \frac{-4(\alpha p_F)^2\tau}{\nu
v_F^2}\int_0^\infty\frac{d\omega}{2\pi}\big[\frac{\partial}{\partial
\omega}(\omega \coth\frac{\omega}{2k_BT})\big]\nonumber\\
&&=\frac{2(\alpha p_F)^2\tau}{\pi\nu
v_F^2}(2k_BT-E_F\coth\frac{E_F}{2k_BT}),
\end{eqnarray}
where $\tan^{-1}(\omega\tau)$ is replaced by $\pi/2$ for
$\omega\tau\gg 1$ in the ballistic regime
and $E_F$ is in the place of the upper limit of the integral.
In the low temperature regime $k_BT\ll E_F$,
the second term approaches a constant independent of the temperature,
so the first term manifests the temperature effect in the contribution of
the electron-electron interaction to the spin-relaxation time.

When the total spin density $\mathbf{S}$ is spatially homogeneous
and parallel to the $\ell$th-axis of the coordinate frame, the
contribution of $F_\ell(S_x, S_y, S_z)$ vanishes, namely
$F_\ell(S_x, S_y, S_z)=0$. The diffusion equations for spin
components $S_\ell$ can be simplified, for example,
\begin{eqnarray}
\partial_tS_x=-\frac{1}{\tau_s'}S_x+\frac{1}{\tau_{xx}^e}S_x
=-\frac{1}{\tau^s_{xx}}S_x,
\end{eqnarray}
where
$\tau^s_{xx}=\tau_s'/(1-\displaystyle\frac{\tau_s'}{\tau^e_{xx}})$.
Therefore, the spin-relaxation times can be determined by $\tau_s'$
and $\tau_{\ell\ell}^e$, consequently,
\begin{eqnarray}\label{zxt}
&&\tau^s_{xx}=\frac{\tau_s'}{1-\displaystyle\frac{\tau_s'}{\tau^e_{xx}}},\
\tau^s_{yy}=\frac{\tau_s'}{1-\displaystyle\frac{\tau_s'}{\tau^e_{yy}}},\nonumber\\
&&(\tau^s_{zz})^{-1}=(\tau^s_{xx})^{-1}+(\tau^s_{yy})^{-1}.
\end{eqnarray}
We can see that the total spin decays exponentially when
$0<\tau_s'/\tau^e_{\ell\ell}<1$. In terms of the explicit forms of
the characteristic times $\tau_{\ell\ell}^e$ in the ballistic
regime, the spin-relaxation times involving the effect of the
electron-electron interaction take the following forms
\begin{eqnarray}\label{time4}
\tau^s_{xx}=\tau^s_{yy}=2\tau^s_{zz}
=\displaystyle\frac{\tau_s'}{1-(\frac{T}{T_F}-\frac{1}{2})},
   \, T\tau\gg 1,
\end{eqnarray}
where $T_F=E_F/k_B$ is the Fermi temperature. It is worthwhile to
indicate that there exists an obvious enhancement of the
spin-relaxation time with increment of the temperature in the
ballistic regime. The increasing amplitude of the spin-relaxation
time depends on the ratio of the temperature to the Fermi
temperature. In conclusion, an obvious enhancement of the
spin-relaxation time can be induced by the electron-electron
interaction in the ballistic regime for systems under consideration.

\section{Summary}\label{sec:summary}

In the above, we presented a theoretical study of the influence of
electron-electron interactions on the spin dynamics for 2DEGs with
Rashba spin-orbit coupling. We employed the path-integral approach
and the quasiclassical Green's function to deal with the
electron-electron interaction. With the help of the auxiliary Bose
field, the electron-electron interaction was decoupled via the
Hubbard-Stratonovich transformation. Then one is able to derive the
Elienberger equation by using the Green's function after the
transformation. Through tedious calculation, we further derived the
spin and charge diffusion equations, from which the spin-relaxation
time can be given explicitly.
We analyzed the influence of the
electron-electron interaction on the spin-relaxation time in the
ballistic regime and found an obvious enhancement of the
spin-relaxation time with the increment of the temperature $T$.
The
increasing amplitude of the spin-relaxation time depends on the
ratio of the temperature to the Fermi temperature.
The electron-electron interaction changes
the wave vector $\mathbf{k}$ and hence results in the variation of
the spin precession vector.
This exhibits that the electron-electron
interaction plays an important role in the spin relaxation of
electrons when the D'yakonov-Perel spin relaxation mechanism
dominates. It is expected to be helpful for understanding the spin
dynamics of 2DEGs with spin-orbit couplings and electron-electron
interactions. Our formulation can also be extend to
the case of bulk inversion asymmetry, namely the additional
Dresselhaus term~\cite{Dresselhaus} with
$\mathbf{b}=\beta(p_x,-p_y)+\gamma(p_xp_y^2-p_yp_x^2)$.

\acknowledgments
The work was supported by NSFC Grant No. 10674117
and partially by
PCSIRT Grant No. IRT0754.

\appendix
\section{Explicit forms}\label{sec:explicit}

The explicit form of $\delta g^K_0$ is
\begin{widetext}
\begin{eqnarray}\label{dgk1}
\delta g^K_0(t_1,t_2;\mathbf{n},\mathbf{r})&=&-i\int
dt_\theta\big[\phi_1(\mathbf{r}_1,t_1-t_\theta)-\phi_1(\mathbf{r}_1,t_2-t_\theta)\big]
\int\frac{d\theta'}{2\pi}\Gamma_{\rho}(t_\theta,\mathbf{n},\mathbf{n}';\mathbf{r},\mathbf{r}_1)\nonumber\\
&&\times\Big\{\langle
g^K_0(t_1-t_{\theta},t_2-t_{\theta};\mathbf{n}_1,\mathbf{r})\rangle_{\Phi,
\mathbf{n}_1}
+2\mathbf{n}'\cdot\langle\mathbf{n}_1g^K_0(t_1-t_{\theta},t_2-t_{\theta};\mathbf{n}_1,\mathbf{r})\rangle_{\Phi, \mathbf{n}_1}\Big\}\nonumber\\
&&+\int \frac{d\theta'}{2\pi}\frac{d\theta''}{2\pi}\int d^2r_1
dt_{\theta}\Gamma_{\rho}(t_\theta,\mathbf{n},\mathbf{n}';\mathbf{r},\mathbf{r}_1)
\Big\{2i\phi_2(\mathbf{r}_1,t_1-t_{\theta})\delta(t_1-t_2)\nonumber\\
&&+\frac{i}{\tau}\big[\langle\
\Gamma_{\rho}(t_4-t_3,\mathbf{n}'',\mathbf{n}_1;\mathbf{r_2},\mathbf{r}_1)\
\rangle_{\mathbf{n}_1}
-\Gamma_{\rho}(t_4-t_3,\mathbf{n}'',\mathbf{n};\mathbf{r_2},\mathbf{r}_1)\big]\phi_2(\mathbf{r}_2,t_4)\nonumber\\
&&\times\big[\langle\
g^K_0(t_1-t_{\theta},t_3;\mathbf{n}_1,\mathbf{r})\ \rangle _{\Phi,
\mathbf{n}_1}
\langle\ g^K_0(t_3,t_2-t_{\theta};\mathbf{n}_1,\mathbf{r})\ \rangle_{\Phi, \mathbf{n}_1}\nonumber\\
&&+\langle\
\mathbf{g}^{}_K(t_1-t_{\theta},t_3;\mathbf{n}_1,\mathbf{r})\
\rangle_{\Phi, \mathbf{n}_1}\cdot \langle\
\mathbf{g}^{}_K(t_3,t_2-t_{\theta};\mathbf{n}_1,\mathbf{r})\
\rangle_{\Phi, \mathbf{n}_1}\big]\Big\}.
\end{eqnarray}
The explicit expression of $\delta \mathbf{g}^K$ is
\begin{eqnarray}\label{dgk2}
Q(t_1,t_2;\mathbf{n},\mathbf{r})&=&-i\int dt_\theta
\big[\phi_1(\mathbf{r}_1,t_1-t_\theta)-\phi_1(\mathbf{r}_1,t_2-t_\theta)\big]
\int\frac{d\theta'}{2\pi}
\Gamma_s(t_\theta,\mathbf{n},\mathbf{n}';\mathbf{r},\mathbf{r}_1)\nonumber\\
&\times&\Big\{\langle\
L_k(t_1-t_\theta,t_2-t_\theta;\mathbf{n}_1,\mathbf{r})\
\rangle_{\Phi, \mathbf{n}_1}
+2\mathbf{n}'\cdot\langle\mathbf{n}_1L_k(t_1-t_{\theta},t_2-t_{\theta};\mathbf{n}_1,
\mathbf{r})\rangle_{\Phi, \mathbf{n}_1}\Big\}\nonumber\\
&+&\frac{2i}{\tau}\int
\frac{d\theta'}{2\pi}\frac{d\theta''}{2\pi}\int d^2r_1
dt_{\theta}\Gamma_{s}(t_\theta,\mathbf{n},\mathbf{n}';\mathbf{r},\mathbf{r}_1)
\langle\ g^K_0(t_1-t_\theta,t_3;
\mathbf{n}_1,\mathbf{r})\ \rangle_{\Phi, \mathbf{n}_1}\nonumber\\
&\times&\Big\{\big[\langle\
\Gamma_{\rho}(t_4-t_3,\mathbf{n}'',\mathbf{n}_1;\mathbf{r_2},
\mathbf{r}_1)\
\rangle_{\mathbf{n}_1}-\Gamma_{\rho}(t_4-t_3,\mathbf{n}'',
\mathbf{n}_1;\mathbf{r_2},\mathbf{r}_1)\big]\nonumber\\
&\times&\phi_2(\mathbf{r}_4,t_2) \langle\
L_k(t_3,t_2-t_{\theta};\mathbf{n}_1, \mathbf{r})\ \rangle_{\Phi,
\mathbf{n}_1}\Big\}.
\end{eqnarray}
The inelastic collision integral reads
\begin{eqnarray}\label{c1}
&&C_{in}\big\{\langle
g^K\rangle_\Phi\big\}(t,\epsilon;\mathbf{n},\mathbf{r})\nonumber\\
&=&\frac{i}{2\tau} \int
d^2r_1d^2r_2\int\frac{d\omega}{2\pi}\big[D^R(\omega;\mathbf{r},\mathbf{r}_2)
-D^A(\omega;\mathbf{r}_2,\mathbf{r})\big]\nonumber\\
&&\times\big[\langle\ \Gamma_{\rho}(\omega;\mathbf{r_2},
\mathbf{r}_1)\ \rangle\langle\ \Gamma_{\rho}(-\omega;\mathbf{r},
\mathbf{r}_1)\ \rangle-\langle\ \Gamma_{\rho}(-\omega;\mathbf{r},
\mathbf{r}_1)\Gamma_{\rho}(\omega;\mathbf{r}_2,
\mathbf{r}_1)\ \rangle\big]\nonumber\\
&&\times\big[\langle\
g^K_0(t,\epsilon-\omega;\mathbf{n},\mathbf{r})\rangle_{\Phi,
\mathbf{n}}\ \langle g^K_0(t,\epsilon;\mathbf{n},\mathbf{r})\
\rangle_{\Phi, \mathbf{n}}+\langle\
\mathbf{g}^K(t,\epsilon-\omega;\mathbf{n},\mathbf{r})\
\rangle_{\Phi, \mathbf{n}}
\cdot\langle\ \mathbf{g}^K(t,\epsilon;\mathbf{n},\mathbf{r})\ \rangle_{\Phi, \mathbf{n}}\big]\nonumber\\
&&-\frac{i}{2}\int
d^2r_1\int\frac{d\omega}{2\pi}D^K(\omega;\mathbf{r},\mathbf{r}_1)\sigma_m
\big[\langle\ \Gamma_{s}(-\omega;\mathbf{r}, \mathbf{r}_1)\
\rangle+\langle\ \Gamma_{s}(\omega;\mathbf{r}_1,\mathbf{r})\
\rangle\big] \big[\langle\ L_k(t,\epsilon;\mathbf{n},\mathbf{r})\
\rangle_{\Phi, \mathbf{n}}-\langle\
L_k(t,\epsilon-\omega;\mathbf{n},
\mathbf{r})\ \rangle_{\Phi, \mathbf{n}}\big]\nonumber\\
&&+\frac{i}{\tau}\int
d^2r_1d^2r_2\int\frac{d\omega}{2\pi}\big[D^R(\omega;\mathbf{r},\mathbf{r}_2)
-D^A(\omega;\mathbf{r}_2,\mathbf{r})\big]\sigma_m\nonumber\\
&&\times\big[\langle\Gamma_{s}(-\omega;\mathbf{r}_2,
\mathbf{r}_1)\rangle\langle\ \Gamma_{\rho}(-\omega;\mathbf{r},
\mathbf{r}_1)\rangle-\langle\Gamma_{s}(-\omega;\mathbf{r},\mathbf{r}_1)\Gamma_{\rho}(\omega;\mathbf{r}_2,
\mathbf{r}_1)\rangle\big]\langle
g^K_0(t,\epsilon-\omega;\mathbf{n},\mathbf{r})\rangle_{\Phi,
\mathbf{n}}\langle
L_k(t,\epsilon;\mathbf{n},\mathbf{r})\rangle_{\Phi, \mathbf{n}},
\end{eqnarray}
\end{widetext}
where
\begin{eqnarray}\label{cin2}
&&C_{in}\big\{\langle
g^K\rangle_\Phi\big\}(t_1,t_2;\mathbf{n},\mathbf{r})\nonumber\\
&=&\int \frac{d \epsilon}{2\pi}C_{in}\big\{\langle
g^K\rangle_\Phi\big\}\big(\frac{t_1+t_2}{2},\epsilon;\mathbf{n},\mathbf{r}\big)e^{i\epsilon(t_2-t_1)},
\end{eqnarray}
and $\langle\Gamma_{\rho\ (s)}\rangle$ means angular averaging
defined in Eq.~(\ref{diff}), the matrix
$\sigma_m=(\sigma_x,\sigma_y,\sigma_z)$ and the temporal
transformation of the Green's function has been used due to a much
faster dependence on the difference $t_1-t_2$ than on the $t_1+t_2$
\begin{eqnarray}\label{bh}
g^K(t_1,t_2;\mathbf{n},\mathbf{r})=\int \frac{d
\epsilon}{2\pi}g^K\big(\frac{t_1+t_2}{2},\epsilon;\mathbf{n},\mathbf{r}\big)e^{i\epsilon(t_2-t_1)},
\end{eqnarray}
the propagators of auxiliary fields have the same transformation.
The elastic collision integral can be written as
\begin{widetext}
\begin{eqnarray}\label{c2}
C_{el}\big\{\langle
g^K\rangle_\Phi\big\}(t,\epsilon;\mathbf{n},\mathbf{r})&=&
-\frac{2}{\tau}\mathbf{n}_\imath\langle\ \mathbf{n}_\imath
g^K(t,\epsilon;\mathbf{n},\mathbf{r})\ \rangle_{\Phi,
\mathbf{n}}-\frac{2}{\tau}\int \frac{d\omega}{2\pi}\mathbf{n}_\imath
R_1^{\imath\jmath}(\omega)\langle\ g^K_0(t,\epsilon-\omega;\mathbf{n},\mathbf{r})\ \rangle_{\Phi, \mathbf{n}}\nonumber\\
&&\times\langle\ \mathbf{n}_\jmath
g^K(t,\epsilon;\mathbf{n},\mathbf{r})\ \rangle_{\Phi, \mathbf{n}}
+\frac{i}{\tau}\int \frac{d\omega}{2\pi}\mathbf{n}_\imath
R_2^{\imath\jmath}(\omega)\langle\ \mathbf{n}_\jmath
g^K(t,\epsilon-\omega;\mathbf{n},\mathbf{r})\ \rangle_{\Phi, \mathbf{n}}\langle\ g^K(t,\epsilon;\mathbf{n},\mathbf{r})\
\rangle_{\Phi, \mathbf{n}}\nonumber\\
&&-\frac{i}{\tau}\int \frac{d\omega}{2\pi}\mathbf{n}_\imath
\big(R_2^{\imath\jmath}(\omega)\big)^\star \langle\
g^K(t,\epsilon;\mathbf{n},\mathbf{r})\ \rangle_{\Phi,
\mathbf{n}}\langle\ \mathbf{n}_\jmath
g^K(t,\epsilon-\omega;\mathbf{n},\mathbf{r})\ \rangle_{\Phi, \mathbf{n}}\nonumber\\
&&+\frac{i}{\tau}\int \frac{d\omega}{2\pi}\mathbf{n}_\imath
\Big[\sigma_m R_3^{\imath\jmath}(\omega)\langle\
L_k(t,\epsilon-\omega;\mathbf{n},\mathbf{r})\ \rangle_{\Phi,
\mathbf{n}}\langle\ \mathbf{n}_\jmath
g^K(t,\epsilon;\mathbf{n},\mathbf{r})\ \rangle_{\Phi, \mathbf{n}}\nonumber\\
&&-\langle\ \mathbf{n}_\jmath g^K(t,\epsilon;\mathbf{n},\mathbf{r})\
\rangle_{\Phi, \mathbf{n}}\
\sigma_m\big(R_3^{\imath\jmath}(\omega)\big)^\star \langle\
L_k(t,\epsilon-\omega;\mathbf{n},\mathbf{r})\ \rangle_{\Phi,
\mathbf{n}}\Big].
\end{eqnarray}
\end{widetext}
Where $\mathbf{n}_{\imath(\jmath)}$ refers to the $\imath(\jmath)$
component of the unit vector $\mathbf{n}$, with $\imath,\jmath=x,y$
and the kernels
$R_1^{\imath\jmath}(\omega)-R_3^{\imath\jmath}(\omega)$ in
Eq.~(\ref{c2}) are defined by
\begin{eqnarray}
&&R_1^{\imath\jmath}(\omega)=\mathrm{Im}\int\frac{d^2q}{(2\pi)^2}D^R(\omega;\mathbf{q})
\big\{\langle\ \Gamma_{\rho}(\mathbf{n},\omega;\mathbf{q})\mathbf{n}_{\jmath}\ \rangle\nonumber\\
&&\times\langle\ \mathbf{n}_{\imath}\Gamma_{\rho}\ \rangle
-\frac{1}{2}\delta_{\imath,\jmath}\big(\langle\
\Gamma_{\rho}\rangle\langle\Gamma_{\rho}\ \rangle -\langle\
\Gamma_{\rho}\Gamma_{\rho}\ \rangle\big)\big\},
\end{eqnarray}
\begin{eqnarray}
&&R_2^{\imath\jmath}(\omega)=\int\frac{d^2q}{(2\pi)^2}D^R(\omega;\mathbf{q})
\big\{\langle\ \Gamma_{\rho}\ \rangle\langle\ \mathbf{n}_{\imath}\Gamma_{\rho}\mathbf{n}_{\jmath}\ \rangle\nonumber\\
&&-\langle\
\Gamma_{\rho}\mathbf{n}_{\imath}\Gamma_{\rho}\mathbf{n}_{\jmath}\
\rangle -\langle\ \Gamma_{\rho}\mathbf{n}_{\imath}\ \rangle\langle\
\Gamma_{\rho}\mathbf{n}_{\jmath}\ \rangle\big\},
\end{eqnarray}
\begin{eqnarray}
&&R_3^{\imath\jmath}(\omega)=\int\frac{d^2q}{(2\pi)^2}D^R(\omega;\mathbf{q})
\big\{\langle\ \Gamma_{\rho}\mathbf{n}_{\jmath}\ \rangle\langle\ \mathbf{n}_{\imath}\Gamma_{s}\ \rangle\nonumber\\
&&-\frac{1}{2}\delta_{\imath,\jmath}\big(\langle\ \Gamma_{\rho}\
\rangle\langle\ \Gamma_{s}\ \rangle -\langle\
\Gamma_{\rho}\Gamma_{s}\ \rangle\big)\big\},
\end{eqnarray}
where we have introduced the notation
\begin{eqnarray}\label{diff}
\langle f\Gamma_{\rho(s)}h\rangle= \int\frac{d\theta
d\theta'}{(2\pi)^2}f(\mathbf{n})\Gamma_{\rho(s)}
(\mathbf{n},\mathbf{n}';\omega,\mathbf{q})h(\mathbf{n}').
\end{eqnarray}

\section{Calculation of the kernels $R_i^{\imath\jmath}$}\label{sec:calculation}
According to the definition of the diffusion propagator
$\Gamma_{\rho}$ in Eq.~(\ref{gm1}), we can obtain
\begin{eqnarray}
\Gamma_{\rho}(\mathbf{n},\mathbf{n}';\omega,\mathbf{q})&=&
2\pi\delta(\mathbf{n}-\mathbf{n}') \Gamma_{\rho
0}(\mathbf{n};\omega,\mathbf{q})\nonumber\\&+&\Gamma_{\rho
0}(\mathbf{n};\omega,\mathbf{q})\Gamma_{\rho
0}(\mathbf{n}';\omega,\mathbf{q})\displaystyle\frac{1}{\tau-\frac{1}{\Upsilon}},\nonumber\\
\end{eqnarray}
where
\begin{eqnarray}
\Gamma_{\rho
0}(\mathbf{n};\omega,\mathbf{q})&=&\frac{1}{-i\omega+iv_F\mathbf{n}\cdot
\mathbf{q}+\frac{1}{\tau}}\nonumber\\
&=&\frac{1}{-i\omega+iv_Fq\cos(\phi-\phi_q)+\frac{1}{\tau}},\nonumber\\
\Upsilon&=&\sqrt{(-i\omega+\frac{1}{\tau})^2+v_F^2q^2},
\end{eqnarray}
with $\phi_q$ being the angle between the wave vector $\mathbf{q}$
and the $x$-axis. In terms of the explicit form of the diffusion
propagator, one can obtain
\begin{eqnarray}
\langle\Gamma_{\rho}
\rangle&=&\frac{\tau}{\Upsilon\tau-1},\nonumber\\
\langle\Gamma_{\rho}n_x \rangle&=&\langle n_x\Gamma_{\rho}
\rangle=\frac{\tau}{iv_Fq(\Upsilon\tau-1)}(\Upsilon+
i\omega-\frac{1}{\tau})\cos\phi_q,\nonumber\\
\langle\Gamma_{\rho}n_y \rangle&=&\langle n_y\Gamma_{\rho}
\rangle=\frac{\tau}{iv_Fq(\Upsilon\tau-1)}(\Upsilon+
i\omega-\frac{1}{\tau})\sin\phi_q,\nonumber\\
\langle\Gamma_{\rho}\Gamma_{\rho}
\rangle&=&\frac{-i\omega+\frac{1}{\tau}}{\Upsilon(\Upsilon-\frac{1}{\tau})^2},\nonumber\\
\langle\Gamma_{\rho}n_x\Gamma_{\rho}
\rangle&=&\frac{1}{\Upsilon}\sin^2\phi_q\nonumber\\
&&-\frac{\Upsilon}{v_F^2q^2
(\Upsilon\tau-1)}(1-\frac{-i\omega+\frac{1}{\tau}}{\Upsilon})^2\cos^2\phi_q,\nonumber\\
\langle\Gamma_{\rho}n_y\Gamma_{\rho}
\rangle&=&\frac{1}{\Upsilon}\cos^2\phi_q\nonumber\\
&&-\frac{\Upsilon}{v_F^2q^2
(\Upsilon\tau-1)}(1-\frac{-i\omega+\frac{1}{\tau}}{\Upsilon})^2\sin^2\phi_q,\nonumber\\
\langle\Gamma_{\rho}n_x\Gamma_{\rho}
n_x\rangle&=&\frac{\tau}{\Upsilon\tau-1}\big(\frac{-i\omega
+\frac{1}{\tau}}{\Upsilon^2}\sin^2\phi_q\nonumber\\
&&\hspace{1cm}-\frac{\Upsilon-(i\omega+\frac{1}
{\tau})}{\Upsilon^2(\Upsilon\tau-1)}\cos^2\phi_q\big),\nonumber\\
\langle\Gamma_{\rho}n_y\Gamma_{\rho}
n_y\rangle&=&\frac{\tau}{\Upsilon\tau-1}\big(\frac{-i\omega
+\frac{1}{\tau}}{\Upsilon^2}\cos^2\phi_q\nonumber\\
&&\hspace{1cm}-\frac{\Upsilon-(i\omega+\frac{1}
{\tau})}{\Upsilon^2(\Upsilon\tau-1)}\sin^2\phi_q\big).\nonumber\\
\end{eqnarray}
Utilizing above formulas, we find that the kernels
$R_1^{\imath\jmath}(\omega)$ and $R_2^{\imath\jmath}(\omega)$ are
diagonal, $R_i^{\imath\jmath}=\delta_{\imath\jmath}R_i$, which can
be written as
\begin{eqnarray}\label{r1}
R_1(\omega)&=&-\mathrm{Im}\int_0^\infty
\frac{qdq}{4\pi}D^R(\omega;q)\Big\{\frac{1}{v_F^2q^2}
\big(\frac{\Upsilon+i\omega-\frac{1}{\tau}}{\Upsilon-\frac{1}{\tau}}\big)^2\nonumber\\
&&\hspace{3cm}+\frac{\Upsilon+i\omega-\frac{1}{\tau}}{\Upsilon(\Upsilon-\frac{1}{\tau})^2}\Big\},\nonumber\\
\mathrm{Im}R_2(\omega)&=&\mathrm{Im}\int_0^\infty
\frac{qdq}{4\pi}D^R(\omega;q)\Big\{\frac{1}{v_F^2q^2}
\frac{[\Upsilon+i\omega-\frac{1}{\tau}]^2}{\Upsilon(\Upsilon-\frac{1}{\tau})}\nonumber\\
&&\hspace{3cm}+\frac{\Upsilon+i\omega-\frac{1}{\tau}}{\Upsilon(\Upsilon-\frac{1}{\tau})^2}\Big\}.
\end{eqnarray}

It is not difficult to calculate the concrete forms of the electron
polarization operators from Eq.~(\ref{polar1}), for example,
\begin{eqnarray}
\Pi^R(\omega,q)=\nu[1+\frac{i\omega}{\Upsilon-\frac{1}{\tau}}].
\end{eqnarray}
Substituting the polarization operator into Eq.~(\ref{polar2}), we
obtain the propagator of the Bose fields, i.e.,
\begin{eqnarray}
D^R(\omega,q)&=&\frac{D_0^R}{1-D_0^R\Pi^R}=\frac{-2\pi
e^2/q}{1+\frac{2\pi
e^2}{q}\nu[1+\frac{i\omega}{\Upsilon-\frac{1}{\tau}}]}\nonumber\\
&\approx &
-\frac{1}{\Pi^R}=-\frac{1}{\nu}\frac{\Upsilon-\frac{1}{\tau}}{\Upsilon-\frac{1}{\tau}-i\omega},
\end{eqnarray}
where the approximation in the second line corresponds to the
unitary limit associating with larger distances than the screening
radius.

We obtain the concrete expressions of the kernels $R_1$ and $R_2$ in
the ballistic regime $T\tau\gg 1$,
\begin{eqnarray}\label{ballistic}
R_1(T\tau\gg 1)\propto \frac{1}{8\pi\nu
v_F^2}(\frac{3\pi}{2}+\tan^{-1}\omega\tau-\frac{2\omega\tau}{1+\omega^2\tau^2}), \nonumber\\
\mathrm{Im}R_2(T\tau\gg 1)\propto -\frac{1}{8\pi\nu
v_F^2}(\frac{3\pi}{2}+\tan^{-1}\omega\tau-\frac{2\omega\tau}{1+\omega^2\tau^2}).\nonumber\\
\end{eqnarray}


\begin{thebibliography}{99}
\bibitem{onAverage}
Note that the average over a variable means to ingtegrate the
variable out simply, while the average over a functional $\phi$
alway implies an additional meassure of Feymann path-integral,
$e^{-iS[\phi]}$, in the functional intergration.
\end{thebibliography}
\end{document}